\documentclass[conference]{IEEEtran}
\IEEEoverridecommandlockouts
\usepackage{cite}
\usepackage{threeparttable}
\usepackage[table,xcdraw]{xcolor}
\usepackage{array}
\usepackage{multirow}
\usepackage{amsmath,amssymb,amsfonts}
\usepackage{graphicx}
\usepackage{textcomp}
\usepackage{svg}
\usepackage{booktabs}
\usepackage{multirow}
\usepackage[caption=false, font=footnotesize]{subfig}
\usepackage{mathtools}

\usepackage{algorithm}
\usepackage{algpseudocode}

\makeatletter
\newcommand*\titleheader[1]{\gdef\@titleheader{#1}}
\AtBeginDocument{%
  \let\st@red@title\@title
  \def\@title{%
    \bgroup\normalfont\small\centering\@titleheader\par\egroup
    \vskip0.2em\st@red@title}
}
\makeatother

\usepackage{xcolor}
\usepackage{todonotes}
\usepackage[a4paper, total={184mm,239mm}]{geometry}
\def\BibTeX{{\rm B\kern-.05em{\sc i\kern-.025em b}\kern-.08em
    T\kern-.1667em\lower.7ex\hbox{E}\kern-.125emX}}

\titleheader{THIS IS A PREPRINT - Accepted at 
Design, Automation and Test in Europe Conference (DATE) 2024}
\title{MATADOR: Automated System-on-Chip Tsetlin Machine Design Generation for Edge Applications}
\begin{document}

\author{
        Tousif Rahman$^{*\dagger\ddagger}$, Gang Mao$^{*\dagger\ddagger}$, Sidharth Maheshwari$^{\diamond\ddagger}$, Rishad Shafik$^{*}$, Alex Yakovlev$^{*}$ \\\small{$^{*}$Microsystems Research Group, Newcastle University, Newcastle upon Tyne, UK. $^{\diamond}$ IIT Jammu, Jammu, India}  \\\small{$\{$s.rahman, g.mao2, rishad.shafik, alex.yakovlev$\}$@newcastle.ac.uk, sidharth.maheshwari@iitjammu.ac.in}\\\small{$^\dagger$Indicates equal contribution, $\ddagger$ Indicates corresponding authors.}
}
\maketitle
\thispagestyle{plain}
\pagestyle{plain}
\begin{abstract}


System-on-Chip Field-Programmable Gate Arrays (SoC-FPGAs) offer significant throughput gains for machine learning (ML) edge inference applications via the design of co-processor accelerator systems. However, the design effort for training and translating ML models into SoC-FPGA solutions can be substantial and requires specialist knowledge aware trade-offs between model performance, power consumption, latency and resource utilization. Contrary to other ML algorithms, Tsetlin Machine (TM) performs classification by forming logic proposition between boolean actions from the Tsetlin Automata (the learning elements) and boolean input features. A trained TM model, usually, exhibits high sparsity and considerable overlapping of these logic propositions both within and among the classes. The model, thus, can be translated to RTL-level design using a miniscule number of \texttt{AND} and \texttt{NOT} gates. This paper presents MATADOR, an automated boolean-to-silicon tool with GUI interface capable of implementing optimized accelerator design of the TM model onto SoC-FPGA for inference at the edge. It offers automation of the full development pipeline: model training, system level design generation, design verification and deployment. It makes use of the logic sharing that ensues from propositional overlap and creates a compact design by effectively utilizing the TM model's sparsity. MATADOR accelerator designs are shown to be up to 13.4x faster, up to 7x more resource frugal and up to 2x more power efficient when compared to the state-of-the-art Quantized and Binary Deep Neural Network implementations. 
\end{abstract}

\section{Introduction}
\label{Sec:introduction}

Inference at the edge for Machine Learning (ML) classification tasks presents notable advantages of increased data security, reduced network-bandwidth and lower energy requirements in data movement when compared to cloud off-loading based approaches~\cite{FINN_R, Gobieski2019}. The bottleneck, however, stems from the restrictive resources available at the edge~\cite{FINN, DAC_16}. To deliver high throughput within the limited compute, memory and power budget, the ML algorithms employ techniques such as compression and quantization~\cite{REDRESS,FINN, Deng2020}. These software savings can then be further amplified into hardware accelerator solutions that leverage compute pipelining and parallelism for greater throughput and energy efficiency~\cite{FINN, DAC_16}.

Recent approaches with quantization of Deep Neural Network (DNN) models down to 1 or 2-bit weight and activation representation have demonstrated the possibility of using DNNs at the edge~\cite{FINN_R,ConvNet}. These quantized models permit reduced memory usage enabling network parameters to remain on-chip, thus, avoiding data movement latency. With quantized weights and activation functions, the computation can be performed using integer operations reducing the resource, computation time, and power requirements~\cite{FINN, FINN_R}. There are, however, some drawbacks and limitations to accelerating and fully automating the offline training, design generation and edge deployment for DNN based designs. Firstly, there is an accuracy degradation as floating-point parameters are quantized to fixed-bit(s) representation. Secondly, even with quantization, considerable design space explorations of DNN architectures are required to find an optimum trade-off between performance, network parameters and energy efficiency~\cite{Deng2020}. This remains a key challenge as it requires substantial automation efforts in hardware-software co-design space~\cite{DAC_16} making automated ML hardware generation and integration into a system level solutions complex~\cite{DAC_16, Zhang2021}.

\begin{figure}[t]
    \centering
    \includegraphics[width =0.96\linewidth]{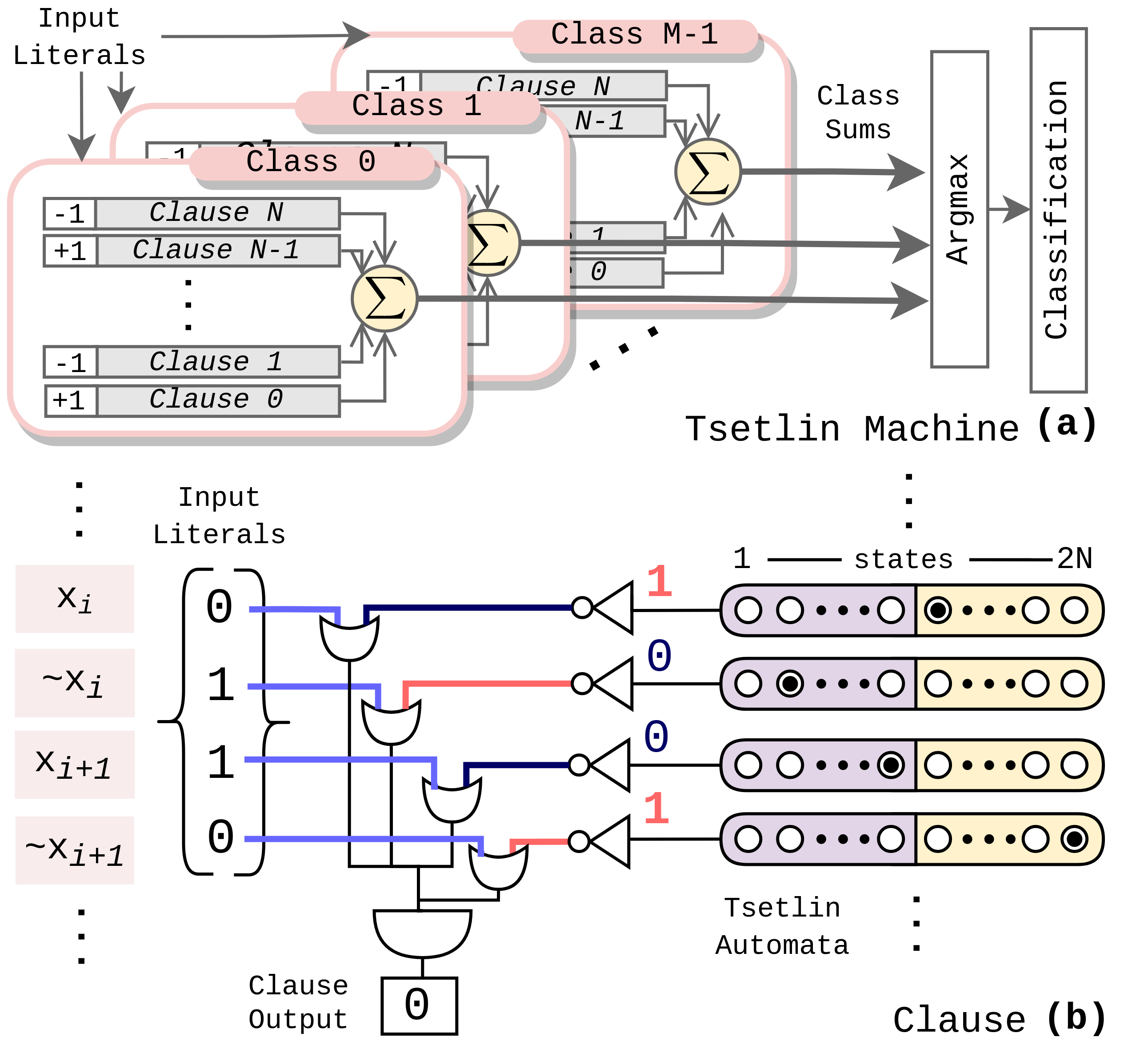}
    \vspace{-3.5mm}
    \caption{Visualization of the Tsetlin Machine and its main components. (a) shows the clause voting mechanism that results in class sums and classification. (b) internals of the clause with the learning element Tsetlin Automata and propositional logic that knits it with input literals (equivalent of features).} 
    \label{fig:TM_overview}
    \vspace{-6.8mm}
\end{figure}

This paper looks beyond DNNs to address the aforementioned challenges through exploring end-to-end automated hardware accelerator design of a recently proposed ML algorithm called the Tsetlin Machine (TM), as shown in Fig.~\ref{fig:TM_overview}. The TM classifies through voting mechanism of the clause outputs that are summed together in accordance with their polarity as shown in Fig~\ref{fig:TM_overview}(a), where the polarity changes alternatively between [+1,-1]. The clause is the ML compute engine of TM that consists of two crucial components: Tsetlin automata and propositional logic. Tsetlin automata is the main learning element that interprets the input data through changes in the state value while training, as shown in Fig~\ref{fig:TM_overview}(b). The final states obtained upon training are translated into boolean actions of \texttt{0} or \texttt{1} resulting in a long boolean sequence termed as the TM model. The propositional logic is formed through interlinking the input literals with their respective boolean actions to generate a clause output as shown in Fig~\ref{fig:TM_overview}(b). Here, each feature $x_{i}$ will produce two literals: $x_{i}$ and its inverse $\sim x_{i}$. For further details on the TM training and inference, the interested readers are encouraged to read~\cite{REDRESS,granmo2021tsetlin}. The intrinsic logic arithmetic and a limited design space exploration, with few hyperparameters~\cite{REDRESS}, involved in TM qualifies it for edge translation using a System-on-Chip Field-Programmable Gate Array (SoC-FPGA).

This paper presents MATADOR, a boolean-to-silicon automation tool with a GUI interface that is capable of training Tsetlin Machines with user-guided hyperparameters, generating RTL-level design for the model and mapping it to SoC-FPGA architectures. MATADOR realizes the TM inference with Look-up-Table friendly \texttt{AND} and \texttt{NOT} operations emanating from the boolean actions of the automata used to create logic propositions, as shown in Fig~\ref{fig:TM_overview}(b). MATADOR can, essentially, translate the entire TM model to a single combinational circuit. However, due to the bandwidth constraint between the processor and the programmable logic, and to avoid timing violations, it automatically splits it into smaller units, depending on the number of input literals, each calculating a partial clause output. 

MATADOR is the first to present a full system design automation approach, exploiting the algorithm-centric sparsity of Tsetlin Machine models. The fundamental design concept for the MATADOR accelerator takes inspiration from the streaming architectures presented in ~\cite{FINN, ConvNet}. They build custom architectures for Binary Neural Network (BNN) and  Quantized Neural Network (QNN) based topologies, respectively, with dedicated compute engines for each layer (or group of layers as in ~\cite{ConvNet}). In this paper, we demonstrate that MATADOR accelerator designs can be better in resource frugality and power efficiency in comparison to these state-of-the-art alternatives.

\section{Boolean-to-Silicon}
\label{Sec:TM}
The Tsetlin machine~\cite{granmo2021tsetlin} represents a significant departure from the floating-point arithmetic-based machine learning to a logic-based approach. At its core, it employs clauses containing learning automata, as finite-state machines, to acquire logical patterns in relation to the input dataset. These logical clauses form comprehensive descriptions of the learning problem. Consequently, the Tsetlin machine introduces the concept of logical interpretable learning, wherein both the learned model and the learning process are easily comprehensible and explainable~\cite{Dropped_Clause_TM}. As a result, it reduces the level of expertise required to efficiently apply machine learning techniques across various domains, in tandem, the parallelism in logic based inference can be further exploited in the hardware.

Recent developments with TM have explored applications such as Internet-of-Things~\cite{Hotmobile, Bakar2023, REDRESS} and estimating battery parameters in electric vehicles~\cite{evtm}. This has lead to a democratic development of novel TM architectures and training methods such as convolutional TM~\cite{granmo2019convolutional}  with superior classification results, regression convolutional TM~\cite{Abeyrathna_Granmo_Goodwin_2021}, and Coalesed TM~\cite{glimsdal2021coalesced} with small memory footprint while training. TM architecture and hyperparameter search paradigms are proposed in \cite{REDRESS, mileage, olga}. On-chip TM training needs to consider area, power, frequency and high-throughput of pseudo random numbers, being a stochastic process, as discussed in~\cite{trim, cyclostationary, spng}. Energy-efficient and fast FPGA TM inference is proposed in  \cite{RoySoC, Svein}, however, they have used small datasets viz. 2D Noisy XOR and IRIS~\cite{misc_iris_53}. MATADOR extends inference with TMs to larger edge application datasets presented in Section~\ref{Sec:eval}.

\begin{figure}[t]
    \centering
    \includegraphics[width =\linewidth]{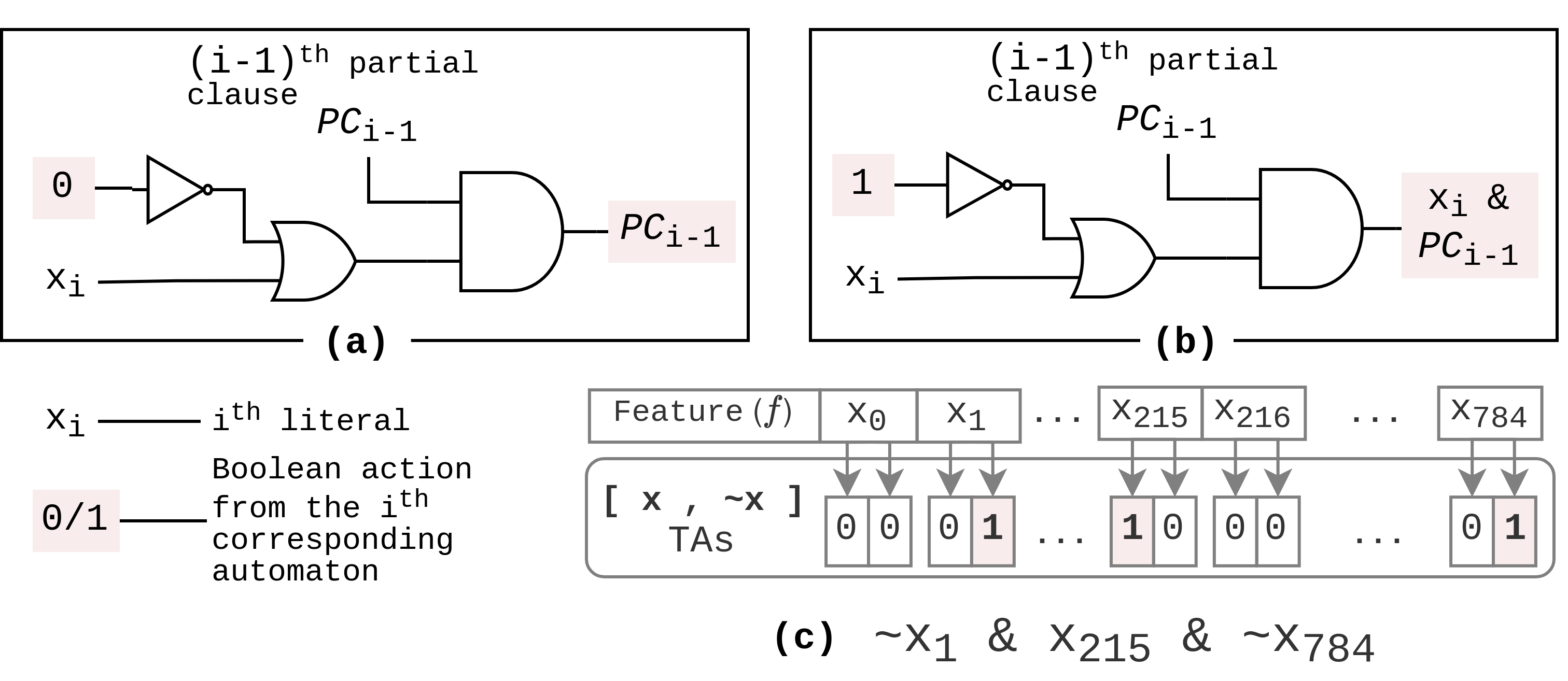}
    \vspace{-6.9mm}
    \caption{Following Fig.~\ref{fig:TM_overview}, here we visualize the outcomes of partial clauses (\texttt{\textit{PC}} via the propositional logic for different boolean actions of the automaton. (a,b) if boolean action is \texttt{0} then the corresponding literal can be excluded from clause computation on the other hand it needs to be included if the action is \texttt{1}. (c) the include-excludes result in boolean expression for the incoming input.} 
    \label{fig:clause_comp}
\end{figure}

\begin{figure}[t]
    \centering
    \vspace{-3.5mm}
    \includegraphics[width =\linewidth]{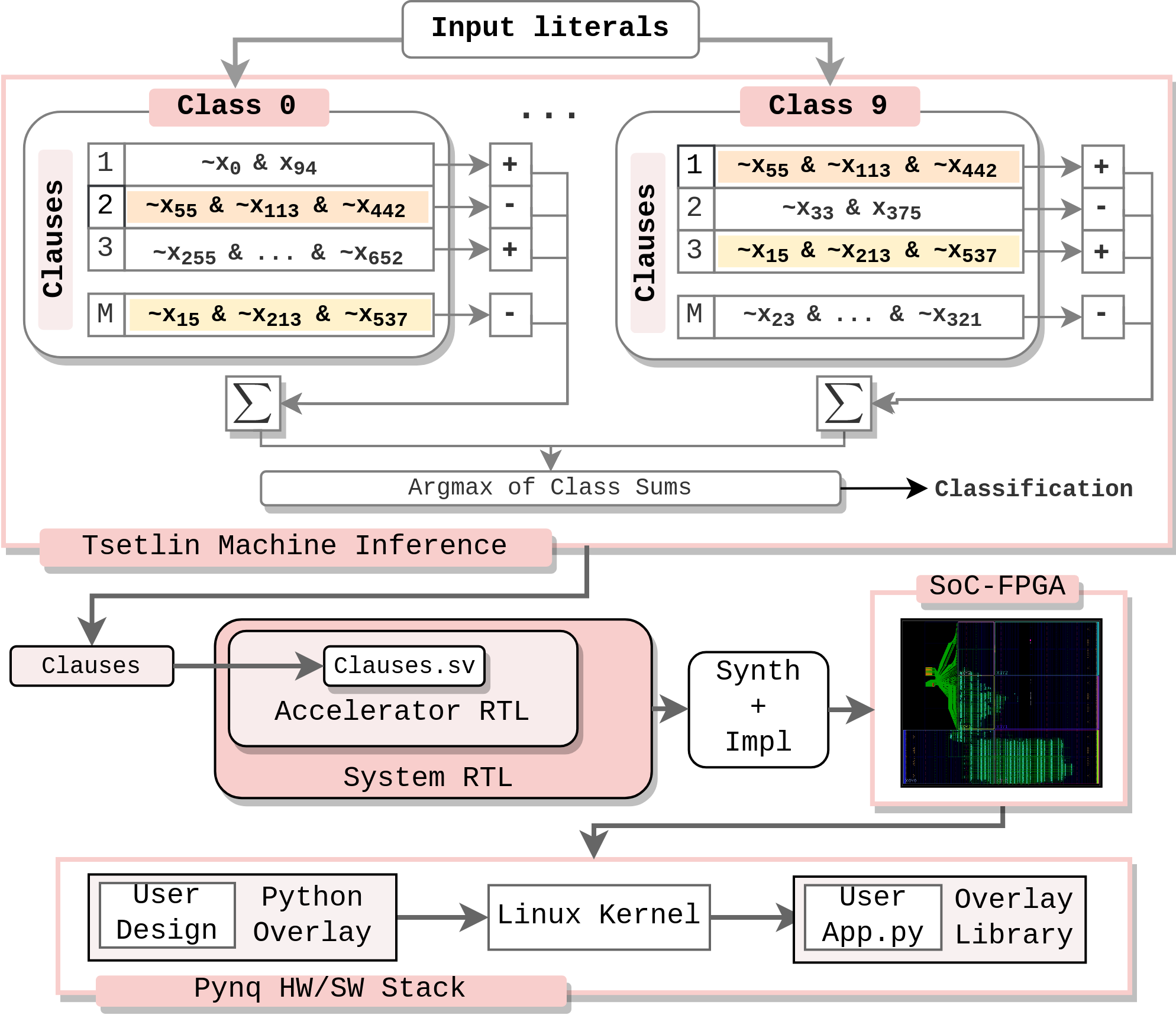}
    \vspace{-5.5mm}
    \caption{Boolean-to-Silicon Overview: The boolean expression from the include-exclude decisions of the automata, are often found to be similar within various clauses, and classes, as highlighted in the Tsetlin Machine Inference box. Logic sharing can be used to shrink the combination circuit arising from the TM model. The clauses are the basis from which the SoC-FPGA system is built.} 
    \label{fig:TM_inf}
    \vspace{-5mm}
\end{figure}

\begin{figure*}[t]
    \centering
    \includegraphics[width =0.93\textwidth]{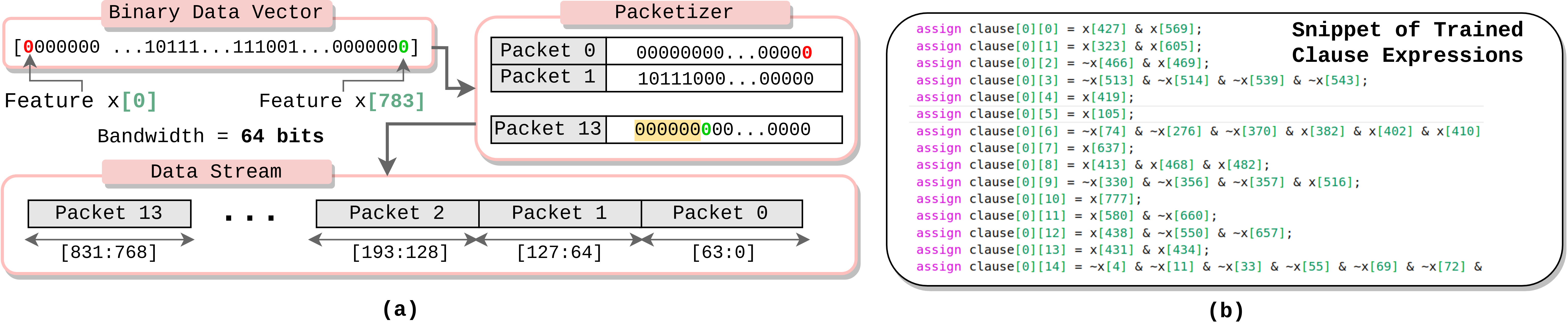}
    \vspace{-3.5mm}
    \caption{(a): Packetization of a MNIST datapoint in the processor of the SoC-FPGA. (b): A snippet of the full clause expressions generated by MATADOR after training the Tsetlin Machine on MNIST. The clauses are a 2D array of dimensions \texttt{[number of classes][number of clauses]}.} 
    \label{fig:datapacket_clause_Diagram}
    \vspace{-2mm}
\end{figure*}
\begin{figure*}[t]
    \centering
    \includegraphics[width =0.93\textwidth]{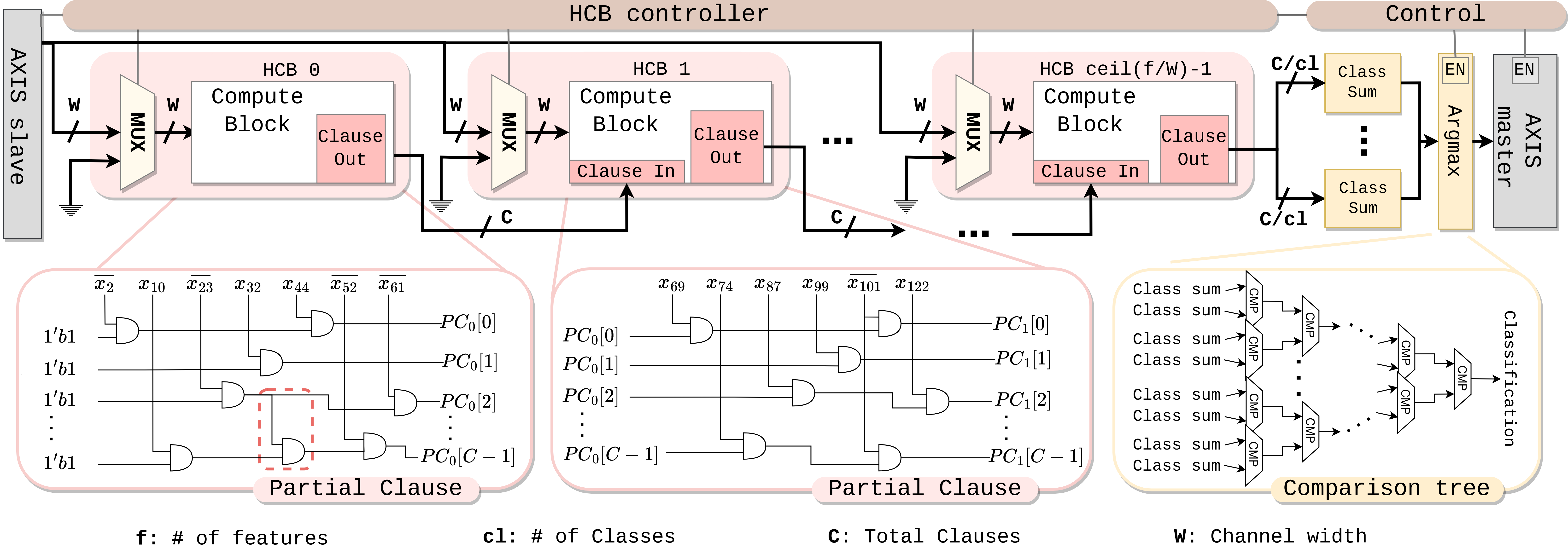}
    \vspace{-2.5mm}
    \caption{Block diagram of the generated MATADOR inference accelerator architecture.} 
    \label{fig:Accel_Block_Diagram}
    \vspace{-4mm}
\end{figure*}
Fig.~\ref{fig:clause_comp} showcases how a TM model can be translated into a combinational circuit. If the boolean action of the automaton is \texttt{0} then the corresponding literal will never affect the clause output and, thus, the related \texttt{NOT} and \texttt{AND} gates, shown in Fig.~\ref{fig:clause_comp}(a), can be excluded from the combinational circuit. Similarly, the boolean action \texttt{1} mandates that the literal needs to included in the circuit as it will determine the clause output, shown in Fig.~\ref{fig:clause_comp}(b). The clause output is an accumulation of partial clause outputs. Fig.~\ref{fig:clause_comp}(c) shows the translation of include(\texttt{1})-exclude(\texttt{0}) decision into a boolean expression. Fig.~\ref{fig:TM_inf} presents a crucial aspect of Tsetlin Machines that enables a resource frugal, performance and power efficient {boolean-to-silicon} transition of TM models. From extensive experimentation, we have empirically found that TM models exhibit extremely high sparsity in the occurrence of includes, and significant sharing of boolean expressions among the clauses within the class as well as among the classes. This observation is pivotal, leading to condensed and compact designs via sparsity and logic sharing. This along with an automated RTL to SoC-FPGA implementation, making use of the open-source \textit{Pynq} Hardware/Software stack, makes MATADOR a true {boolean-to-silicon} framework.


\section{Accelerator Architecture}
\label{Sec:Hardware implementation}
The MATADOR design methodology takes its namesake from achieving the fastest possible inference time for a given model with respect to bandwidth - MATADOR: autoMated dATa bAndwidth Driven lOgic based infeRence. To that end, all the major compute units are designed to complete operations within one clock cycle. This section presents the design decisions that efficiently use the bandwidth constraints, the sparsity of includes in the trained TM and how this can all be incorporated into the compute units of accelerator. 


\textbf{Bandwidth Driven Data Partitioning:} The generated accelerator architecture works with streamed input data where the system latency depends on the effective use of channel bandwidth. The input is, therefore, sent to the accelerator from the processor in packets. Fig.~\ref{fig:datapacket_clause_Diagram} shows the packetization and AXI4-stream protocol based streaming of the data.

Fig.~\ref{fig:datapacket_clause_Diagram}(a) shows the packetization of an MNIST example with binary input vector of 784 bits. Assuming the channel bandwidth between the processor and FPGA chip to be 64 bits, 13 packets are needed to send each datapoint. On the processor side the \textit{Packetizer} orders the data from the least significant bit (seen in red) and adds the appropriate zero padding to the last packet (see the yellow highlight after the most significant bit seen in green in Packet 13). For reference, a snippet of clause expressions of MNIST trained TM model is shown in Fig.~\ref{fig:datapacket_clause_Diagram}(b).    

\textbf{Hard Coded Clause Blocks (HCBs):} The staggered arrival of the input data posits a crucial design challenge - the organization of the clause compute unit in order to process the arriving time-multiplexed data-packets. Attempting to buffer all the packets of one datapoint before computing the full clause output in a single clock cycle leads to extensive sequential chains. As verified through experimentation, synthesizing these chains leads to many critical paths, timing violations and prevents MATADOR running at higher frequencies.

\begin{figure*}[ht]
    \centering
    \includegraphics[width =0.92\textwidth]{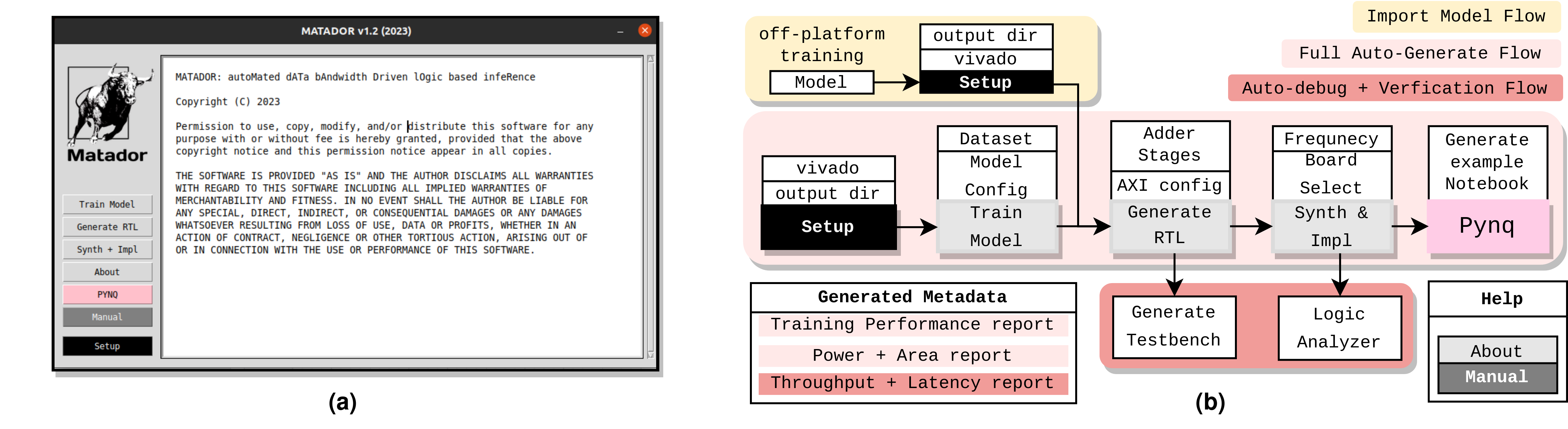}
    \vspace{-3mm}
    \caption{Matador Design Flow. (a) shows the GUI used to guide the user through design space exploration and accelerator implementation. (b) shows the flow for how the design is generated - the main flow is shown in the pink. Each block is stacked with its respective dependencies. The yellow flow shows how TM models trained outside the MATADOR tool can be imported in. The dark pink flow shows the automated design verification.} 
    \label{fig:Tool}
    \vspace{-3mm}
\end{figure*}
\begin{figure}[ht]
    \centering
    \includegraphics[width =\linewidth]{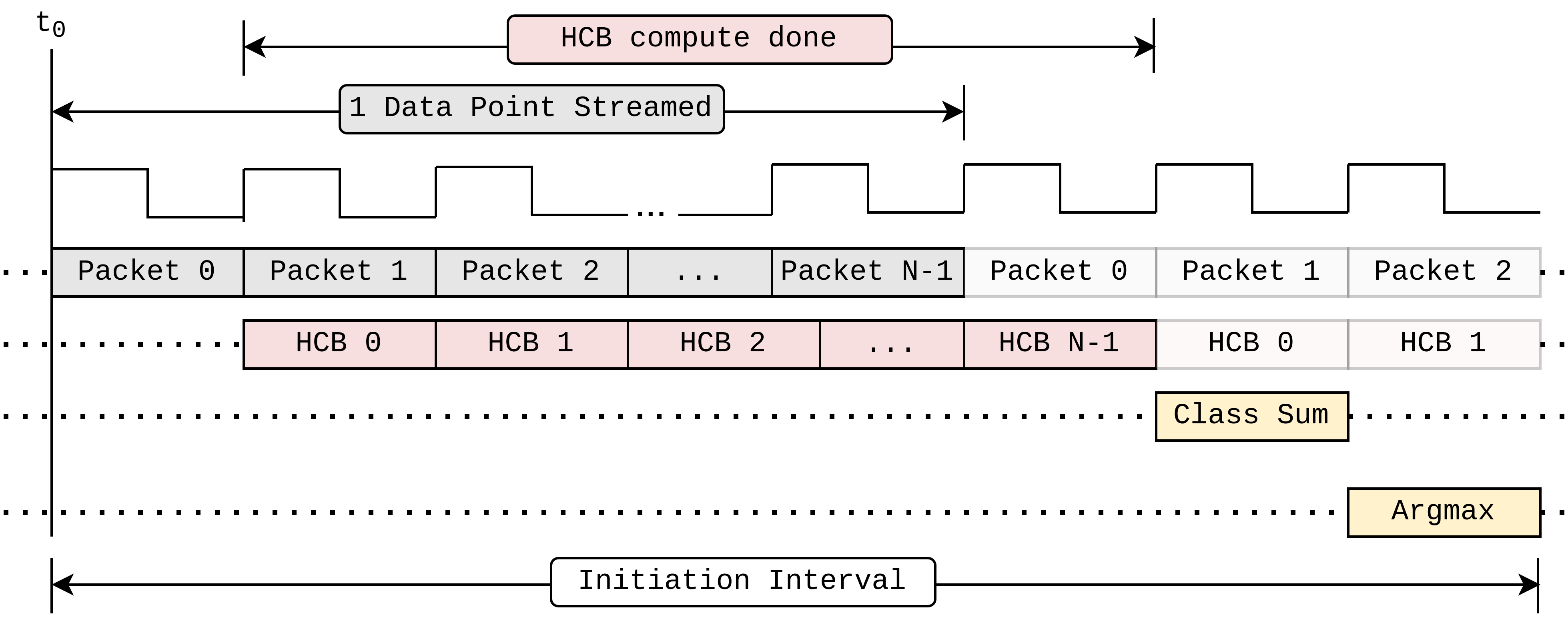}
    \vspace{-5.5mm}
    \caption{Timing diagram depicting the routing of each data packet to its respective HCB and the pipelining opportunities in the subsequent Class Sum and Argmax stages.} 
    \label{fig:timing}
    \vspace{-5.5mm}
\end{figure}

Instead, the MATADOR inference architecture splits the clauses into partial clauses encoded as Hard Coded Clause Blocks (HCB)s, as shown in Fig.~\ref{fig:Accel_Block_Diagram}. The clause expressions are divided into the respective indexes of the data packets, i.e. the first data packet of 64 bits ([\texttt{63:0}]) will interact with all the clauses (indicated as \texttt{C}) in all the classes containing the first 64 corresponding include decisions, as shown in Fig.~\ref{fig:Accel_Block_Diagram}. The next packet will be routed to clauses that contain the corresponding next 64 ([\texttt{127:64}]) include decisions in relation to the packet. The multiplexer in HCB determines when to accept data from the stream. Fig.~\ref{fig:Accel_Block_Diagram} shows two such partial clause units \texttt{HCB 0} and \texttt{HCB 1} where \texttt{HCB 0} initializes all the clauses in the system to  \texttt{1'b1}, as it starts with the assumption that all clause outputs are 1.




The gate-level description of the partial clause, shown in Fig.~\ref{fig:Accel_Block_Diagram}, visualizes logic sharing using the highlighted dashed box in partial clause cut-away of \texttt{HCB 0}. Logic sharing occurs in both intra- and inter-units. Unlike the streaming architectures used in ~\cite{FINN}, where each compute block is seen as its own core, MATADOR combines the full design into one core to exploit logic sharing across all the HCB blocks, that is optimized during synthesis and routing.

The partial clause outputs ($PC_{i}$) stored in the \texttt{Clause Out} part of HCB is passed to next HCB, as \texttt{Clause In}, upon the arrival of the next packet. Therefore, the clause expressions continue to get evaluated until the final expressions are formed by the last HCB (the total number of HCBs is equal to the total number of packets needed for one datapoint). The control of the HCB blocks is performed by a dedicated HCB controller.    


\textbf{Class Sum and Argmax Computation:} The class sum blocks tally the final clause votes, for each class, in according to the polarity of the clauses. Positive and negative polarity clause votes are accumulated separately and summed in the end to obtain final class sum for each class. Therefore, each class sum block has $2 \times$\texttt{cl} adders, where \texttt{cl} is number of classes. The MATADOR tool allows users to pipeline these adders. 

MATADOR employs binary tree comparison, illustrated in Fig~\ref{fig:Accel_Block_Diagram}, to compute Argmax to find the largest class sum and reports the associated index as the classification. This design is parameterized, adopting a comparison tree with $2^{\lceil \log(cl) \rceil}$ inputs. Any classes beyond the actual count are assigned the minimum value at the input stage.

\textbf{System Overview and Latency:} The inference architecture is orchestrated from a dedicated control unit. This unit is used to handle the AXI-stream transactions and offer reset, stall, compute and idle functionalities to the architecture. At the core of the MATADOR design methodology is the concept of being bandwidth driven, the throughput of the MATADOR inference engines are close to the rate at which datapoints are transmitted between the processor and the FPGA fabric. The best model size and performance for the given application can then be determined from this property. 

Fig.~\ref{fig:timing} examines the latency more thoroughly. It shows the initiation interval for the first datapoint that is streamed in. Each packet is routed to its respective HCB until the final full clause is computed, only then the class sum can be computed. This part of the design prevents the pipelining, but the class sum and argmax stages may be pipelined. Subsequent datapoints can be inferred at rate equal to the number of packets required to send the full datapoint. The simplicity of the design means TM models can be translated to RTL without high-level synthesis (HLS) abstractions.    

{\renewcommand{\arraystretch}{1.4}
\begin{table*}
\caption{Table showing how MATADOR inference accelerators compare vs the closest comparable state-of-the-art BNNs and QNNs.}
\vspace{-3 mm}
\label{tab:main_results}
\resizebox{\textwidth}{!}{%
\begin{tabular}{lrrrrrrrrrrrrr|}
\hline
\multicolumn{1}{|c|}{} &
  \multicolumn{8}{c|}{\cellcolor[HTML]{D9D9D9}\textbf{Resource Usage}} &
  \multicolumn{1}{c|}{} &
  \multicolumn{1}{c|}{} &
  \multicolumn{1}{c|}{} &
  \multicolumn{1}{c|}{} &
  \multicolumn{1}{c|}{} \\ \cline{2-9}
\multicolumn{1}{|c|}{\multirow{-2}{*}{Models}} &
  \multicolumn{1}{c|}{\cellcolor[HTML]{D9D9D9}LUTs} &
  \multicolumn{1}{c|}{\cellcolor[HTML]{D9D9D9}Slice Registers} &
  \multicolumn{1}{c|}{\cellcolor[HTML]{D9D9D9}F7 Mux} &
  \multicolumn{1}{c|}{\cellcolor[HTML]{D9D9D9}F8 Mux} &
  \multicolumn{1}{c|}{\cellcolor[HTML]{D9D9D9}Slice} &
  \multicolumn{1}{c|}{\cellcolor[HTML]{D9D9D9}LUT as logic} &
  \multicolumn{1}{c|}{\cellcolor[HTML]{D9D9D9}LUT as mem} &
  \multicolumn{1}{c|}{\cellcolor[HTML]{D9D9D9}BRAM} &
  \multicolumn{1}{c|}{\multirow{-2}{*}{Test Acc (\%)}} &
  \multicolumn{1}{c|}{\multirow{-2}{*}{Total Pwr (W)}} &
  \multicolumn{1}{c|}{\multirow{-2}{*}{Dyn Pwr (W)}} &
  \multicolumn{1}{c|}{\multirow{-2}{*}{Latency (us)}} &
  \multicolumn{1}{c|}{\multirow{-2}{*}{Throughput (inf/s)}} \\ \hline
\multicolumn{14}{|c|}{\textbf{MNIST}} \\ \hline
\rowcolor[HTML]{F4CCCC} 
\multicolumn{1}{|l|}{\cellcolor[HTML]{F4CCCC}BNN-r-ref} &
  \multicolumn{1}{r|}{\cellcolor[HTML]{F4CCCC}5636} &
  \multicolumn{1}{c|}{\cellcolor[HTML]{F4CCCC}-} &
  \multicolumn{1}{c|}{\cellcolor[HTML]{F4CCCC}-} &
  \multicolumn{1}{c|}{\cellcolor[HTML]{F4CCCC}-} &
  \multicolumn{1}{c|}{\cellcolor[HTML]{F4CCCC}-} &
  \multicolumn{1}{c|}{\cellcolor[HTML]{F4CCCC}-} &
  \multicolumn{1}{c|}{\cellcolor[HTML]{F4CCCC}-} &
  \multicolumn{1}{r|}{\cellcolor[HTML]{F4CCCC}16} &
  \multicolumn{1}{r|}{\cellcolor[HTML]{F4CCCC}95.83} &
  \multicolumn{1}{r|}{\cellcolor[HTML]{F4CCCC}0.4} &
  \multicolumn{1}{c|}{\cellcolor[HTML]{F4CCCC}-} &
  \multicolumn{1}{r|}{\cellcolor[HTML]{F4CCCC}240} &
  12200 \\
\rowcolor[HTML]{F4CCCC} 
\multicolumn{1}{|l|}{\cellcolor[HTML]{F4CCCC}BNN-f-ref} &
  \multicolumn{1}{r|}{\cellcolor[HTML]{F4CCCC}91131} &
  \multicolumn{1}{c|}{\cellcolor[HTML]{F4CCCC}-} &
  \multicolumn{1}{c|}{\cellcolor[HTML]{F4CCCC}-} &
  \multicolumn{1}{c|}{\cellcolor[HTML]{F4CCCC}-} &
  \multicolumn{1}{c|}{\cellcolor[HTML]{F4CCCC}-} &
  \multicolumn{1}{c|}{\cellcolor[HTML]{F4CCCC}-} &
  \multicolumn{1}{c|}{\cellcolor[HTML]{F4CCCC}-} &
  \multicolumn{1}{r|}{\cellcolor[HTML]{F4CCCC}4.5} &
  \multicolumn{1}{r|}{\cellcolor[HTML]{F4CCCC}95.83} &
  \multicolumn{1}{r|}{\cellcolor[HTML]{F4CCCC}7.3} &
  \multicolumn{1}{c|}{\cellcolor[HTML]{F4CCCC}-} &
  \multicolumn{1}{r|}{\cellcolor[HTML]{F4CCCC}0.31} &
  12361000 \\ \hline
\rowcolor[HTML]{F3F3F3} 
\multicolumn{1}{|l|}{\cellcolor[HTML]{F3F3F3}FINN} &
  \multicolumn{1}{r|}{\cellcolor[HTML]{F3F3F3}11622} &
  \multicolumn{1}{r|}{\cellcolor[HTML]{F3F3F3}17990} &
  \multicolumn{1}{r|}{\cellcolor[HTML]{F3F3F3}172} &
  \multicolumn{1}{r|}{\cellcolor[HTML]{F3F3F3}16} &
  \multicolumn{1}{r|}{\cellcolor[HTML]{F3F3F3}6207} &
  \multicolumn{1}{r|}{\cellcolor[HTML]{F3F3F3}10425} &
  \multicolumn{1}{r|}{\cellcolor[HTML]{F3F3F3}1197} &
  \multicolumn{1}{r|}{\cellcolor[HTML]{F3F3F3}14.5} &
  \multicolumn{1}{r|}{\cellcolor[HTML]{F3F3F3}93.17} &
  \multicolumn{1}{r|}{\cellcolor[HTML]{F3F3F3}1.599} &
  \multicolumn{1}{r|}{\cellcolor[HTML]{F3F3F3}1.458} &
  \multicolumn{1}{r|}{\cellcolor[HTML]{F3F3F3}1.047} &
  954457 \\ \hline
\multicolumn{1}{|l|}{\textbf{MATADOR}} &
  \multicolumn{1}{r|}{\cellcolor[HTML]{FFFFFF}8709} &
  \multicolumn{1}{r|}{\cellcolor[HTML]{FFFFFF}17440} &
  \multicolumn{1}{r|}{\cellcolor[HTML]{FFFFFF}5} &
  \multicolumn{1}{r|}{\cellcolor[HTML]{FFFFFF}0} &
  \multicolumn{1}{r|}{\cellcolor[HTML]{FFFFFF}4186} &
  \multicolumn{1}{r|}{\cellcolor[HTML]{FFFFFF}8516} &
  \multicolumn{1}{r|}{\cellcolor[HTML]{FFFFFF}193} &
  \multicolumn{1}{r|}{\cellcolor[HTML]{FFFFFF}3} &
  \multicolumn{1}{r|}{95.48} &
  \multicolumn{1}{r|}{1.427} &
  \multicolumn{1}{r|}{1.292} &
  \multicolumn{1}{r|}{0.32} &
  \cellcolor[HTML]{FFFFFF}3846153 \\ \hline
\multicolumn{14}{|c|}{\textbf{KWS-6}} \\ \hline
\rowcolor[HTML]{F3F3F3} 
\multicolumn{1}{|l|}{\cellcolor[HTML]{F3F3F3}FINN} &
  \multicolumn{1}{r|}{\cellcolor[HTML]{F3F3F3}42757} &
  \multicolumn{1}{r|}{\cellcolor[HTML]{F3F3F3}45473} &
  \multicolumn{1}{r|}{\cellcolor[HTML]{F3F3F3}206} &
  \multicolumn{1}{r|}{\cellcolor[HTML]{F3F3F3}8} &
  \multicolumn{1}{r|}{\cellcolor[HTML]{F3F3F3}13160} &
  \multicolumn{1}{r|}{\cellcolor[HTML]{F3F3F3}41639} &
  \multicolumn{1}{r|}{\cellcolor[HTML]{F3F3F3}1118} &
  \multicolumn{1}{r|}{\cellcolor[HTML]{F3F3F3}126.5} &
  \multicolumn{1}{r|}{\cellcolor[HTML]{F3F3F3}84.6} &
  \multicolumn{1}{r|}{\cellcolor[HTML]{F3F3F3}3.002} &
  \multicolumn{1}{r|}{\cellcolor[HTML]{F3F3F3}2.796} &
  \multicolumn{1}{r|}{\cellcolor[HTML]{F3F3F3}1.33} &
  750188 \\ \hline
\multicolumn{1}{|l|}{\textbf{MATADOR}} &
  \multicolumn{1}{r|}{\cellcolor[HTML]{FFFFFF}6063} &
  \multicolumn{1}{r|}{\cellcolor[HTML]{FFFFFF}10658} &
  \multicolumn{1}{r|}{\cellcolor[HTML]{FFFFFF}5} &
  \multicolumn{1}{r|}{\cellcolor[HTML]{FFFFFF}0} &
  \multicolumn{1}{r|}{\cellcolor[HTML]{FFFFFF}3609} &
  \multicolumn{1}{r|}{\cellcolor[HTML]{FFFFFF}5878} &
  \multicolumn{1}{r|}{\cellcolor[HTML]{FFFFFF}185} &
  \multicolumn{1}{r|}{\cellcolor[HTML]{FFFFFF}3} &
  \multicolumn{1}{r|}{\cellcolor[HTML]{FFFFFF}87.1} &
  \multicolumn{1}{r|}{\cellcolor[HTML]{FFFFFF}1.422} &
  \multicolumn{1}{r|}{\cellcolor[HTML]{FFFFFF}1.287} &
  \multicolumn{1}{r|}{\cellcolor[HTML]{FFFFFF}0.18} &
  \cellcolor[HTML]{FFFFFF}8333333 \\ \hline
\multicolumn{14}{|c|}{\textbf{CIFAR-2}} \\ \hline
\rowcolor[HTML]{F3F3F3} 
\multicolumn{1}{|l|}{\cellcolor[HTML]{F3F3F3}FINN} &
  \multicolumn{1}{r|}{\cellcolor[HTML]{F3F3F3}23247} &
  \multicolumn{1}{r|}{\cellcolor[HTML]{F3F3F3}25654} &
  \multicolumn{1}{r|}{\cellcolor[HTML]{F3F3F3}162} &
  \multicolumn{1}{r|}{\cellcolor[HTML]{F3F3F3}0} &
  \multicolumn{1}{r|}{\cellcolor[HTML]{F3F3F3}9824} &
  \multicolumn{1}{r|}{\cellcolor[HTML]{F3F3F3}22221} &
  \multicolumn{1}{r|}{\cellcolor[HTML]{F3F3F3}1026} &
  \multicolumn{1}{r|}{\cellcolor[HTML]{F3F3F3}66} &
  \multicolumn{1}{r|}{\cellcolor[HTML]{F3F3F3}81.91} &
  \multicolumn{1}{r|}{\cellcolor[HTML]{F3F3F3}2.206} &
  \multicolumn{1}{r|}{\cellcolor[HTML]{F3F3F3}2.042} &
  \multicolumn{1}{r|}{\cellcolor[HTML]{F3F3F3}0.74} &
  1369879 \\ \hline
\multicolumn{1}{|l|}{\textbf{MATADOR}} &
  \multicolumn{1}{r|}{\cellcolor[HTML]{FFFFFF}3867} &
  \multicolumn{1}{r|}{\cellcolor[HTML]{FFFFFF}33212} &
  \multicolumn{1}{r|}{\cellcolor[HTML]{FFFFFF}5} &
  \multicolumn{1}{l|}{\cellcolor[HTML]{FFFFFF}} &
  \multicolumn{1}{r|}{\cellcolor[HTML]{FFFFFF}6283} &
  \multicolumn{1}{r|}{\cellcolor[HTML]{FFFFFF}3682} &
  \multicolumn{1}{r|}{\cellcolor[HTML]{FFFFFF}185} &
  \multicolumn{1}{r|}{\cellcolor[HTML]{FFFFFF}3} &
  \multicolumn{1}{r|}{\cellcolor[HTML]{FFFFFF}84.8} &
  \multicolumn{1}{r|}{\cellcolor[HTML]{FFFFFF}1.501} &
  \multicolumn{1}{r|}{\cellcolor[HTML]{FFFFFF}1.364} &
  \multicolumn{1}{r|}{\cellcolor[HTML]{FFFFFF}0.38} &
  \cellcolor[HTML]{FFFFFF}3125000 \\ \hline
\multicolumn{14}{|c|}{\textbf{FMNIST}} \\ \hline
\rowcolor[HTML]{F3F3F3} 
\multicolumn{1}{|l|}{\cellcolor[HTML]{F3F3F3}FINN} &
  \multicolumn{1}{r|}{\cellcolor[HTML]{F3F3F3}40002} &
  \multicolumn{1}{r|}{\cellcolor[HTML]{F3F3F3}48901} &
  \multicolumn{1}{r|}{\cellcolor[HTML]{F3F3F3}214} &
  \multicolumn{1}{r|}{\cellcolor[HTML]{F3F3F3}0} &
  \multicolumn{1}{r|}{\cellcolor[HTML]{F3F3F3}13126} &
  \multicolumn{1}{r|}{\cellcolor[HTML]{F3F3F3}38894} &
  \multicolumn{1}{r|}{\cellcolor[HTML]{F3F3F3}1108} &
  \multicolumn{1}{r|}{\cellcolor[HTML]{F3F3F3}131} &
  \multicolumn{1}{r|}{\cellcolor[HTML]{F3F3F3}85.2} &
  \multicolumn{1}{r|}{\cellcolor[HTML]{F3F3F3}2.82} &
  \multicolumn{1}{r|}{\cellcolor[HTML]{F3F3F3}2.622} &
  \multicolumn{1}{r|}{\cellcolor[HTML]{F3F3F3}4.3} &
  232114 \\ \hline
\multicolumn{1}{|l|}{\textbf{MATADOR}} &
  \multicolumn{1}{r|}{\cellcolor[HTML]{FFFFFF}13388} &
  \multicolumn{1}{r|}{\cellcolor[HTML]{FFFFFF}40280} &
  \multicolumn{1}{r|}{\cellcolor[HTML]{FFFFFF}5} &
  \multicolumn{1}{r|}{\cellcolor[HTML]{FFFFFF}0} &
  \multicolumn{1}{r|}{\cellcolor[HTML]{FFFFFF}8289} &
  \multicolumn{1}{r|}{\cellcolor[HTML]{FFFFFF}13195} &
  \multicolumn{1}{r|}{\cellcolor[HTML]{FFFFFF}193} &
  \multicolumn{1}{r|}{\cellcolor[HTML]{FFFFFF}3} &
  \multicolumn{1}{r|}{\cellcolor[HTML]{FFFFFF}87.67} &
  \multicolumn{1}{r|}{\cellcolor[HTML]{FFFFFF}1.501} &
  \multicolumn{1}{r|}{\cellcolor[HTML]{FFFFFF}1.364} &
  \multicolumn{1}{r|}{0.32} &
  \cellcolor[HTML]{FFFFFF}3846153 \\ \hline
\multicolumn{14}{|c|}{\textbf{KMNIST}} \\ \hline
\rowcolor[HTML]{F3F3F3} 
\multicolumn{1}{|l|}{\cellcolor[HTML]{F3F3F3}FINN} &
  \multicolumn{1}{r|}{\cellcolor[HTML]{F3F3F3}40206} &
  \multicolumn{1}{r|}{\cellcolor[HTML]{F3F3F3}49069} &
  \multicolumn{1}{r|}{\cellcolor[HTML]{F3F3F3}186} &
  \multicolumn{1}{r|}{\cellcolor[HTML]{F3F3F3}28} &
  \multicolumn{1}{r|}{\cellcolor[HTML]{F3F3F3}13093} &
  \multicolumn{1}{r|}{\cellcolor[HTML]{F3F3F3}38894} &
  \multicolumn{1}{r|}{\cellcolor[HTML]{F3F3F3}1108} &
  \multicolumn{1}{r|}{\cellcolor[HTML]{F3F3F3}131} &
  \multicolumn{1}{r|}{\cellcolor[HTML]{F3F3F3}89.31} &
  \multicolumn{1}{r|}{\cellcolor[HTML]{F3F3F3}2.695} &
  \multicolumn{1}{r|}{\cellcolor[HTML]{F3F3F3}2.503} &
  \multicolumn{1}{r|}{\cellcolor[HTML]{F3F3F3}3.9} &
  255127 \\ \hline
\multicolumn{1}{|l|}{\textbf{MATADOR}} &
  \multicolumn{1}{r|}{\cellcolor[HTML]{FFFFFF}13911} &
  \multicolumn{1}{r|}{\cellcolor[HTML]{FFFFFF}48539} &
  \multicolumn{1}{r|}{\cellcolor[HTML]{FFFFFF}5} &
  \multicolumn{1}{r|}{\cellcolor[HTML]{FFFFFF}0} &
  \multicolumn{1}{r|}{\cellcolor[HTML]{FFFFFF}10499} &
  \multicolumn{1}{r|}{\cellcolor[HTML]{FFFFFF}13718} &
  \multicolumn{1}{r|}{\cellcolor[HTML]{FFFFFF}193} &
  \multicolumn{1}{r|}{\cellcolor[HTML]{FFFFFF}3} &
  \multicolumn{1}{r|}{\cellcolor[HTML]{FFFFFF}88.6} &
  \multicolumn{1}{r|}{\cellcolor[HTML]{FFFFFF}1.483} &
  \multicolumn{1}{r|}{\cellcolor[HTML]{FFFFFF}1.347} &
  \multicolumn{1}{r|}{0.32} &
  \cellcolor[HTML]{FFFFFF}3846153 \\ \hline
\end{tabular}%
}
\begin{tablenotes}{\footnotesize}
\small{\item[1] The ``-" indicates this result is not reported. Dyn Pwr = Dynamic Power - taken from \textit{Vivado} implementation reports.}
\end{tablenotes}{\footnotesize}
\end{table*}}

\begin{figure*}[t]
    \centering
    \includegraphics[width =\textwidth]{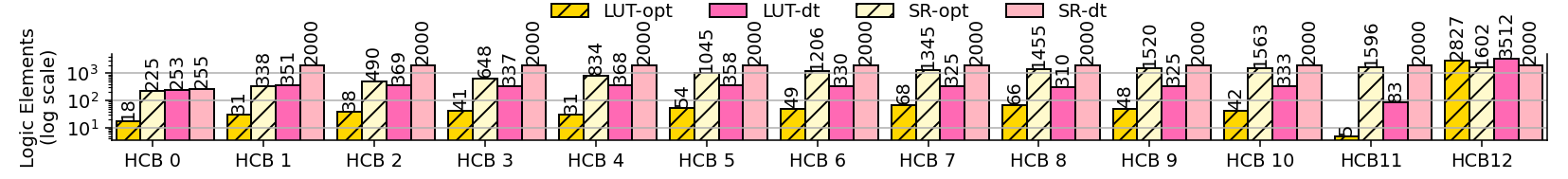}
    \vspace{-7mm}
    \caption{Demonstration of the optimization opportunities from logic sharing using a MNIST model. LUT-opt and SR-opt are the Look-up-Table and Slice Register counts, respectively, post implementation in \textit{Vivado}. LUT-dt and SR-dt are the counts  when the \texttt{DON'T TOUCH} pragma is added to prevent optimizations.} 
    \label{fig:opt}
    \vspace{-5mm}
\end{figure*}

{\renewcommand{\arraystretch}{1.4}{
\begin{table}[]
\caption{Table showing the models used for evaluation}
\vspace{-3 mm}
\label{tab:models}
\resizebox{\linewidth}{!}{%
\begin{tabular}{|l|ll|}
\hline
 &
  \multicolumn{2}{c|}{\textbf{Model}} \\ \cline{2-3} 
\multirow{-2}{*}{\textbf{Dataset}} &
  \multicolumn{1}{c|}{\cellcolor[HTML]{EFEFEF}FINN} &
  \multicolumn{1}{c|}{MATADOR} \\ \hline
 &
  \multicolumn{1}{l|}{\cellcolor[HTML]{F4CCCC}BNN-r/f-ref: 256-256-256 1 bit Weight, Input and Activation Qunatization} &
   \\
\multirow{-2}{*}{MNIST} & \multicolumn{1}{l|}{FINN:784-64-64-64-10 1 bit Weight, Input and Activation Qunatization} & \multirow{-2}{*}{200 Clauses per Class} \\ \hline
KWS6 &
  \multicolumn{1}{l|}{377-512-256-6 1 bit Input, 2 bit Weight and Activation Qunatization} &
  300 Clauses per Class \\ \hline
CIFAR 2 &
  \multicolumn{1}{l|}{1024-256-128-2 1 bit input 1 bit Weight 2 bit Activation Quantization} &
  1000 Clauses per Class \\ \hline
FMNIST &
  \multicolumn{1}{l|}{784-256-256-10 1 bit input 2 bit Weight 2 bit Activation Quantization} &
  500 Claues per Class \\ \hline
KMNIST &
  \multicolumn{1}{l|}{784-256-256-10 1 bit input 2 bit Weight 2 bit Activation Quantization} &
  500 Clauses per Class \\ \hline
\end{tabular}%
}
\begin{tablenotes}
\small{\item[1] MATADOR configuration states the number of clauses per class.}
\vspace{-4.6mm}
\end{tablenotes}
\end{table}

\vspace{-1.2mm}

\section{MATADOR Automation Tool}
This section presents the automation blocks that create the MATADOR framework. From an end-user perspective MATADOR is a GUI interface that can guide them through the accelerator design flow from training the model, generating the RTL design, synthesizing and implementing on a specified SoC-FPGA and creating an example test code to examine throughput metrics with no coding effort as seen in Fig~\ref{fig:Tool}. 

In this respect it offers greater ease of use than its contemporaries like ~\cite{FINN, FINN_R} which are packaged as hardware libraries that require developer level familiarity with training frameworks such as \textit{Brevitas} or \textit{Theano} as well as other third party library dependencies beyond using Xilinx's \textit{Vits HLS} and \textit{Vivado} for the synthesis and implementation flow. MATADOR also requires the same Xilinx tools but no other third party libraries. MATADOR is an open source tool available here: \texttt{https://github.com/nclaes/matador}.

Given that MATADOR makes use of open-source Xilinx IPs, for validating the throughput MATADOR also supports auto-debug functionality through auto-generated testbench and the use of integrated-logic-analyzer (ILA) debug cores into the design so that AXI-stream transactions can be polled. MATADOR also provides a manual used to guide users with no experience with Tsetlin Machines through the automation. MATADOR also offers a sample jupyter notebook where the generated model can be validated for test accuracy and the throughput and latency can be measured. It follows the same throughput measuring procedure as the FINN flow. Throughput and initial latency can be further validated via the auto-debug flow. Given that MATADOR does not require any FPGA BRAM elements auto-debug functionality does not reduce the resource pool for the actual accelerator. 

\section{Evaluation}
\label{Sec:eval}
This section evaluates auto-generated MATADOR implementations against the latest FINN flow as described in ~\cite{FINN_R} with the closest comparable models chosen to examine resource usage, accuracy, power consumption, latency of one datapoint and throughput. The evaluation presents both the best models reported in ~\cite{FINN} and also implements the FINN flow on the same platform as MATADOR for fair like-for-like comparison.

The generated designs are evaluated with five datasets:

\textbf{MNIST}~\cite{mnist}: This has been used as the main benchmarking dataset for the FINN and FINN-R frameworks with well defined models readily available in the FINN codebase repository. \textbf{KMNIST}~\cite{kmnist} is also tested to examine character recognition. \textbf{CIFAR-2}: A 2 class variant derived from the CIFAR-10~\cite{cifar}. CIFAR-2 groups all images into two classes of either animals or vehicles. Used in conjunction with \textbf{FMNIST}~\cite{fmnist} to show possibilities of this flow in classification applications. \textbf{Google Speech Commands (KWS):} A modified version of the dataset is used with 6 chosen keywords: yes, no, up, down, left, right. This is referred to as KWS6, it is used to demonstrate audio classification possibilities.

Table~\ref{tab:main_results} shows the comparison of the auto-generated MATADOR implementations against its closest comparable QNN and BNN implementations via the FINN flow. The BNNs highlighted in pink were taken from ~\cite{FINN}, they were implemented on the Zynq ZC706 SoC running at 200MHz. The remaining FINN and MATADOR accelerators were implemented on Zynq XC7Z020 (Pynq Z1), chosen due to it's more constrained logic resources. These FINN models were ran at 100MHz while all MATADOR designs were ran at optimum frequencies per design between 50MHz-65MHz. The model configurations for all accelerators for each dataset are given in Table~\ref{tab:models}.   

The results indicate that MATADOR is capable of offering a middle ground to the FINN BNN models where it can perform inferences at a faster rate than the resource efficient \textit{BNN-r-ref} but has a lower logic footprint compared to \textit{BNN-f-ref}. The main advantages from MATADOR over the remaining models come from the bandwidth driven inference enabling high throughput and the logic sharing between the HCB blocks. To understand the extent of the optimizations that are made, the MNIST HCBs were passed through the synthesis and implementation flow with \texttt{DON'T TOUCH} pragmas as seen in Fig~\ref{fig:opt}.

\section{Conclusion}
\label{Sec:Conc}
This paper presents the first insights into automated edge inference accelerators for Tsetlin Machines via MATADOR, a boolean-to-silicon tool for training of Tsetlin Machines and their translation into tailored SoC-FPGA accelerators. The benefits of MATADOR based accelerators arise from three interlinked points, firstly, the intrinsic logic proposition forming nature of the TM opens the same opportunities in hardware as binary and quantized DNNs. This allows MATADOR to apply the same heterogeneous streaming architecture design principles found in its closest s.o.t.a. alternatives. Secondly the sparsity in the clause expressions learnt from the TM allows for models to be transitioned to logic constrained SoCs, keeping all the trained model parameters on the FPGA fabric and finally the shared literals or groups of literals between clauses can be exploited by the synthesis tool's logic absorption algorithms to create more resource frugal designs. Further work with MATADOR is concerned with accelerating other TM models for scalability to larger datasets.





\bibliographystyle{IEEEtran}
\bibliography{IEEEabrv,ref}


\end{document}